\documentclass[12pt]{article}

\usepackage{geometry}

\usepackage{times}
\usepackage{amsmath}
\allowdisplaybreaks

\usepackage{amssymb}

\usepackage{physics}
\usepackage{cases}
\usepackage{graphicx}
\usepackage{subcaption}
\usepackage{caption}
\usepackage[hidelinks]{hyperref}

\title{High-Dimensional Bell States: A Paradigm Shift for Quantum Illumination}

\author
{Armanpreet Pannu,$^{1}$ Amr S. Helmy,$^{1}$ Hesham El Gamal$^{2}$\\
\\
\normalsize{$^{1}$The Edward S. Rogers Department of Electrical and Computer Engineering, University of Toronto}\\
\normalsize{$^{2}$School of Electrical and Information Engineering, Faculty of Engineering, University of Sydney}\\
}

\date{}

\begin{document}

\maketitle

\begin{abstract} 
    This paper solves the open problem of characterizing the performance of quantum illumination (QI) with discrete variable states. By devising a novel quantum measurement approach along with meticulous analysis, our investigation demonstrates that, in the limit as $M \rightarrow \infty$, the maximally entangled $M$ mode Bell state achieves optimal performance, matching the two-mode squeezed vacuum in a high-noise regime and exceeding it in low-noise. This result challenges the dominance of continuous variable states in photonic sensing applications and extends the novelty of QI to regimes where no quantum advantage was believed to exist. A closer analysis reveals that this advantage stems from retained entanglement in the transmitted Bell state, a paradigm-shifting discovery since interaction with the environment in optical systems is believed to break entanglement. The complete mathematical analysis of this work provides granular insights into the interaction between photonic systems and environmental noise, motivating further research into discrete variable quantum sensing.
   \end{abstract}

   \section{Introduction}
   Quantum mechanics, with its unique and counter-intuitive phenomena, has laid the foundation for the development of advanced quantum technologies. The hallmark principle underlying these technologies is quantum entanglement, where two systems become interconnected in such a way that the state of one instantly influences the state of the other, regardless of the distance separating them \cite{EPR_original,bells_thm_original_paper}. In photonic applications, entangled states can be categorized into continuous variable (CV) states, such as squeezed states, and discrete variable (DV) states, such as Bell states \cite{Nielsen_Chuang,quantum_info_CV_braunstein_book}.
   
   The use of entanglement to improve target detection in noisy environments was first proposed by Lloyd, who coined the term “quantum illumination” (QI) \cite{lloyd_original_QI}. In his seminal work, Lloyd considered a maximally entangled DV state, more generally known as a high-dimensional Bell state, where one of the entangled photons was sent to probe the target while the other was held as a reference. It was demonstrated that increasing entanglement between the probe and reference photons could help distinguish a back-reflected signal photon from a background noise photon. The result was counter-intuitive because interaction with a noisy environment is typically considered to break entanglement, yet the findings suggested that entanglement provided an advantage even in this entanglement-breaking channels \cite{Sacchi_entaglement_can_enhance}.
   
   Unfortunately, this original work had a couple of key limitations. Firstly, the analysis assumed single-photon operation, which severely limited the parameter space for validity and raised questions about whether the channel model was truly entanglement-breaking. Secondly, the work compared the entangled Bell state to a non-entangled single-photon state, failing to make a comparison with an optimal classical state to claim a true quantum advantage. In fact, a follow-up study \cite{shapiro_and_lloyd_QI} demonstrated that Lloyd’s quantum illumination protocol was actually inferior to an optimal classical scheme with a coherent state transmitter.

   Although the original protocol using DV Bell states was concluded to be ineffective, it sparked extensive follow-up research in QI using CV states. Studies have demonstrated that a quantum advantage of up to 6 dB in the error exponent can be attained over the optimal classical scheme by employing a two-mode squeezed vacuum (TMSV) state \cite{tan_et_al,QI_story_shapiro,QI_review_sorelli_boust}. The maximum quantum advantage is achieved in high-noise environments with low signal power, whereas only a negligible quantum advantage is achievable in low-noise regimes \cite{nair_and_gu,QI_generic_gaussian_Kasra}. It is believed that the TMSV state attains this quantum advantage despite the entanglement seemingly being broken by the environment. This result has led to several studies on optimal receivers \cite{QI_receivers_Guha,optimum_QI_receiver_shapiro,QI_reciever_no_joint_meas_zhuang_2024}, experimental demonstrations \cite{experimental_QI_italian_2013,experimental_QI_shapiro_2013,experimental_QI_shapiro_2015} and follow-up ideas of QI \cite{guha_enhanced_standoff_sensing_2011, ranging_zhuang, quantum_doppler_lidar_2022}.
   
   Apart from Lloyd’s original work \cite{lloyd_original_QI} and the subsequent critique \cite{shapiro_and_lloyd_QI}, there have been no further attempts to study QI with DV states. It is widely believed that CV states are the definitive choice in sensing applications due to the detrimental effect of environmental noise on entanglement. However, DV states, particularly Bell states, play a pervasive role in quantum applications such as computation, teleportation, and communication, substantiating the importance of establishing their role in sensing.
   
   In this work, we tackle the quantum illumination problem using high-dimensional Bell states. We begin by deriving a comprehensive representation of these states after interaction with the target and environment, demonstrating that, in most cases, some degree of entanglement is retained even in a noisy environment. We then devise a novel quantum measurement that leverages this retained entanglement to show that not only are DV states useful in sensing, but optimal. The validity of the analysis extends beyond the assumption of high noise, low signal, and thermal noise establishing the novelty of QI in applications where the TMSV protocol is ineffective. This work challenges the dominance of CV states in sensing, highlighting the untapped potential of discrete variable sensing and strongly motivating further research in this promising area.

   Throughout this manuscript, we use physics notation for bosonic number states. A light beam's state in a particular mode will be represented by a vector denoted with the ket notation $\ket{*}$ in a Hilbert space $\mathcal{H}$. The basis elements of $\mathcal{H}$ are $\ket{n}$, where $\ket{n}$ is a state with exactly $n$ photons. The dual vector to $\ket{\psi}$ is denoted $\bra{\psi}$. For multiple modes, we use vector notation where a vector $\mathbf{v}$ is denoted with bold lettering. The terms $v_i$ will denote the $i$-th component of $\mathbf{v}$, and the corresponding non-bold lettering indicates the component sum $v = \left\lvert\mathbf{v}\right\rvert = v_1 + \ldots + v_M$. We use $\mathbf{v} \leq \mathbf{w}$ to denote component-wise inequality. In the case of multiple modes, $\ket{\mathbf{n}}$ denotes the state with $n_1$ photons in the first mode, $n_2$ in the second mode, etc. We also consider multiple spatial modes (signal, idler, and environment), indicated with a subscript after the $\ket{*}$ or $\bra{*}$ when not evident from context. Lastly, the following shorthand notation $\ket{\mathbf{n},\mathbf{m}}=\ket{\mathbf{m}}\otimes \ket{\mathbf{n}}$ will be used for tensor products of spatial modes.

   \section{Interaction with the environment}
   We start with the typical $M$-mode maximally entangled Bell state with uniform phase across all modes:
   \begin{align}
    \ket{\psi_1} = \frac{1}{\sqrt{M}}\sum_{i=1}^M \ket{\mathbf{e}_i,\mathbf{e}_i}_{I,S} \label{psi_1}
   \end{align}
   Here, $\mathbf{e}_i$ is the standard basis vector of $\mathbb{R}^M$ with all its elements set to $0$ except for the $i$-th element, which is $1$. The idler modes (denoted with $I$) are stored locally, while the signal modes (denoted with $S$) are sent to a target that may or may not be present. The back-reflected signal (if any) is collected after mixing with the background noise state, which we assume to be:
   \begin{align}
       &\rho_{B}=\sum_{\mathbf{N_B}}P_{\mathbf{N_B}} \ket{\mathbf{N_B}}_B \bra{\mathbf{N_B}}_B \label{eq:rho_b}
   \end{align}
   where $P_{\mathbf{N_B}}$ is the probability of having $\mathbf{N_B}$ photons in the background environment. We assume that the photon number is independently and identically distributed (i.i.d.) in each mode with finite mean, so that $P_\mathbf{N_B}=\prod_i p_{N_{B_i}}$. In previous works, $\rho_B$ has typically been taken to be the thermal state in each mode with mean photon number $\mathcal{N}_B$ or $\frac{\mathcal{N}_B}{1-\eta}$, corresponding to the Bose-Einstein photon number distribution for $p_{N_{B_i}}$.
   
   In the case that the target is not present, the background noise from the environment replaces the signal while the idler becomes a completely mixed state. 
   \begin{align}
       \rho_{abs}&=Tr_{S}\Big[\ket{\psi_1}\bra{\psi_1}\Big] \otimes \rho_B \nonumber
       \\&=
       \sum_{i=1}^M\sum_{\mathbf{N_B} }\frac{P_{\mathbf{N_B}}}{M} \ket{\mathbf{e}_i,\mathbf{N_B}}_{I,S} \bra{\mathbf{e}_i,\mathbf{N_B}}_{I,S}
   \end{align}
   When the target is present, the signal beam is mixed with the environment through a beam splitter of reflectivity $0< \eta \leq 1$. We collect one output of this beam splitter while the other is discarded (traced out). In this case, the uncertainty in the environment results in the mixed signal-idler state:
   \begin{align}
       \rho_{pres}=  \sum_{\mathbf{N_B},\mathbf{N_A}} \ket{\tilde{\phi}_{\mathbf{N_A},\mathbf{N_B}}}_{I,S}\bra{\tilde{\phi}_{\mathbf{N_A},\mathbf{N_B}}}_{I,S} \label{rho_pres}
   \end{align}
   where the sum extends over all vectors $\mathbf{N_B}$ and $\mathbf{N_A}$ with non-negative integer components. The state $\ket{\tilde{\phi}_{\mathbf{N_A}, \mathbf{N_B}}}$ corresponds to an unnormalized version of the signal-idler pure state that one would receive if the environment were measured in the pure state $\ket{\mathbf{N_B}}$ before the interaction and ended in the pure state $\ket{\mathbf{N_A}}$ afterward. These states can be explicitly computed as is done in the supplementary materials:
   
   {\footnotesize
   \begin{subnumcases}{\label{eq:tilde_phi}
    \ket{\tilde{\phi}_{\mathbf{N_A},\mathbf{N_B}}}
    =}
    (-1)^{N_A} \frac{\sqrt{\eta P_{\mathbf{N_B}}q^{\mathbf{N_B}}_{\mathbf{N_A}}}}{\sqrt{M}}  \sum_{i=1}^M {\scriptstyle\left(  \sqrt{N_{B_i}-N_{A_i}+1}  - \frac{N_{A_i}}{\sqrt{N_{B_i}-N_{A_i}+1}} \frac{1-\eta}{\eta} \right) } \scriptstyle{\ket{\mathbf{e}_i,\mathbf{N_B}-\mathbf{N_A}+\mathbf{e}_i} } \text{if } \mathbf{N_A}\leq \mathbf{N_B} & \label{eq:tilde_phi_case_1}
       \\
    (-1)^{N_A}\frac{\sqrt{P_{\mathbf{N_B}}q^{\mathbf{N_B}}_{\mathbf{N_A}} }}{\sqrt{M}}       \frac{\sqrt{1-\eta}}{\sqrt{\eta}} \sqrt{N_{B_i}+1} \ket{\mathbf{e}_i,\mathbf{N_B}-\mathbf{N_A}+\mathbf{e}_i} \qquad \substack{ \text{if } N_{A_i}=N_{B_i}+1 \text{ some } i \\ \text{ and } N_{A_j} \leq N_{B_j} \text{ for } j\neq i }&\label{eq:tilde_phi_case_2} 
           \\
    0  \qquad \text{else} \label{eq:tilde_phi_case_3}
   \end{subnumcases}
   }
   Here, $q^{\mathbf{N_B}}_{\mathbf{N_A}} = \prod_i q^{N_{B_i}}_{N_{A_i}}$, where $q^{N_{B_i}}_{N{A_i}}$ is the probability mass function of the binomial distribution with $N_{B_i}$ trials, $\eta$ as the probability of success, and $N_{A_i}$ as the number of successes. Note that despite $(P_{\mathbf{N_B}})\cdot (q^{\mathbf{N_B}}_{\mathbf{N_A}})$ representing a probability distribution on the pair $(\mathbf{N_B}, \mathbf{N_A})$, it does not correspond to the real-world probability of measuring $\mathbf{N_B}$ environment photons before interaction and $\mathbf{N_A}$ afterward. This probability is instead given by the normalization factor $\braket{\tilde{\phi}_{\mathbf{N_A}, \mathbf{N_B}}}$. This discrepancy is what gives rise to fundamentally quantum effects such as the Hong-Ou-Mandel effect \cite{hong_ou_mandel_original}.
   
   Equation (\ref{eq:tilde_phi}) admits a simple intuitive description. If the number of photons in the environment after the interaction is smaller than the number before in all modes, then the entanglement is preserved as in (\ref{eq:tilde_phi_case_1}). This can be interpreted by noting that an observer at the environment after the interaction cannot comment on the signal-idler pair. However, if any one of the modes has an extra photon after the interaction, then an observer at the environment can conclude the signal photon was in that mode. This would collapse the signal-idler photon pair to have been in that mode eliminating the entanglement as in (\ref{eq:tilde_phi_case_2}). All other cases are not possible as the environment after interaction can have at most one extra photon, hence why the state is $0$ in (\ref{eq:tilde_phi_case_3}) corresponding to a $0$ probability.
   
   As noise increases, the probability that all noise photons in a particular mode will remain in the environment significantly decreases, thereby increasing the likelihood that partial entanglement is retained. In QI applications such as secure communication \cite{passive_eavesdropping_QI_shapiro}, the environment is controlled and measurable, allowing for an exact measurement of the anticipated incoming state. However, in general sensing scenarios where the environment is unknown both before and after measurement, the primary challenge becomes to devise a measurement broad enough to detect states $\ket{\tilde{\phi}_{\mathbf{N_A}, \mathbf{N_B}}}$ with enough success while still maintaining some noise resilience.

   \section{The Optimal Measurement}
   
   To construct our proposed quantum measurement, we first note that the states $\ket{\tilde{\phi}_{\mathbf{N_A}, \mathbf{N_B}}}$ are non-orthogonal for a fixed number of collected noise photons $\mathbf{N_C}=\mathbf{N_B}-\mathbf{N_A}$. For each such $\mathbf{N_C}$, there can be an infinite number of corresponding pairs $(\mathbf{N_B},\mathbf{N_A})$, each of which corresponds to a slightly different incoming state $\ket{\tilde{\phi}_{\mathbf{N_A}, \mathbf{N_B}}}$. In general, constructing a measurement that projects over all possible $\ket{\tilde{\phi}_{\mathbf{N_A}, \mathbf{N_B}}}$ will project over the entire Hilbert space, resulting in zero noise resilience.
   
   Instead, we project only onto the simplest state corresponding to each $\mathbf{N_C}$ which is the state where the number of photons in the environment afterward is zero, so $\mathbf{N_C} = \mathbf{N_B}$. This projector state, which we denote as $\ket{\psi_1 + \mathbf{N_C}}$, is simply the normalized version of $\ket{\tilde{\phi}_{\mathbf{0}, \mathbf{N_C}}}$. The overall measurement is then the projective measurement over the subspace spanned by the states $\ket{\psi_1 + \mathbf{N_C}}$. These states form an orthonormal set, allowing us to easily write out the projector:
   \begin{align}
    \hat{P}= \sum_{\mathbf{N_C}} \ket{\psi_1+\mathbf{N_C}}_{I,S}\bra{\psi_1+\mathbf{N_C}}_{I,S} \label{eq:projector}
       \\
    \ket{\psi_1+\mathbf{N_C}}_{I,S}=  \sum_{i=1}^M \frac{\sqrt{N_{C_i}+1}}{\sqrt{N_C+M}}\ket{\mathbf{e}_i,\mathbf{e}_i+\mathbf{N_C}}_{I,S} \label{eq:projection_states}
   \end{align}
   With the projective measurement $\{\hat{P}, 1 - \hat{P}\}$, the two outcomes correspond to the target being present and absent respectively. The probability of a false alarm can be straightforwardly derived (see supplementary material):
   \begin{align}
    P_{FA}&= Tr [\hat{P} \rho_{abs}] \leq \frac{2}{M} \label{p_FA}
   \end{align}
   This gives us the same $O(\frac{1}{M})$ scaling as in Lloyd's original discrete variable QI protocol \cite{lloyd_original_QI}. This bound can be further tightened to $O(\frac{\mathcal{N}_B}{M})$ in the low-noise case, as having a vacuum state in the environment cannot trigger the outcome corresponding to $\hat{P}$. Intuitively, this probability of false alarm arises because, for each fixed number of collected noise photons $\mathbf{N_C}$, the dimensions of the signal and idler Hilbert space scale with $M^2$, while the environment Hilbert space scales with $M$, resulting in a $1/M$ noise resilience. Taking $M \rightarrow \infty$ allows us to infinitely suppress background noise.
   
   However, increasing $M$ comes with the cost of reducing the probability of a successful detection. For the state $\ket{\tilde{\phi}_{\mathbf{N_A}, \mathbf{N_B}}}$ to be picked up by the detector, it must resemble the corresponding $\ket{\psi_1+\mathbf{N_C}}$, which becomes increasingly difficult. While analytically determining the exact probability of successful detection for arbitrary $M$ is challenging, we can determine this probability in the limit as $M \rightarrow \infty$ (see supplementary material):
       \begin{align}
           1-P_{MD}&= Tr [\hat{P} \rho_{pres}] \nobreak 
            \\
            &\geq
            \eta\left(\frac{ 1}{1+(1-\eta)\langle N_B\rangle_{pm} } \right) \label{p_detection}
       \end{align}
   Here, $P_{MD}$ is the probability of missed detection, and $\langle N_B \rangle_{pm}$ is the average number of noise photons per mode. A close look at the derivation in the supplementary materials reveals that the result holds even if the environment is non-thermal. The only required condition is that the starting background state is of the form (\ref{eq:rho_b}) and the noise photon number in each mode be i.i.d with a finite mean. These assumptions are valid for the thermal noise used in previous works \cite{tan_et_al,nair_and_gu}. Consequently, the probability of successful detection in a single shot can be concluded to be at worst $\tilde{\eta}$, given by the expression (\ref{p_detection}), which is slightly lower than the reflectivity of the target $\eta$ for low $\langle N_B \rangle_{pm}$ and substantially lower for large $\langle N_B \rangle_{pm}$.

   \section{Analysis}
\begin{figure*}[]   
   \begin{subfigure}{0.5\textwidth}
           \centering
           \includegraphics[width=0.95\textwidth]{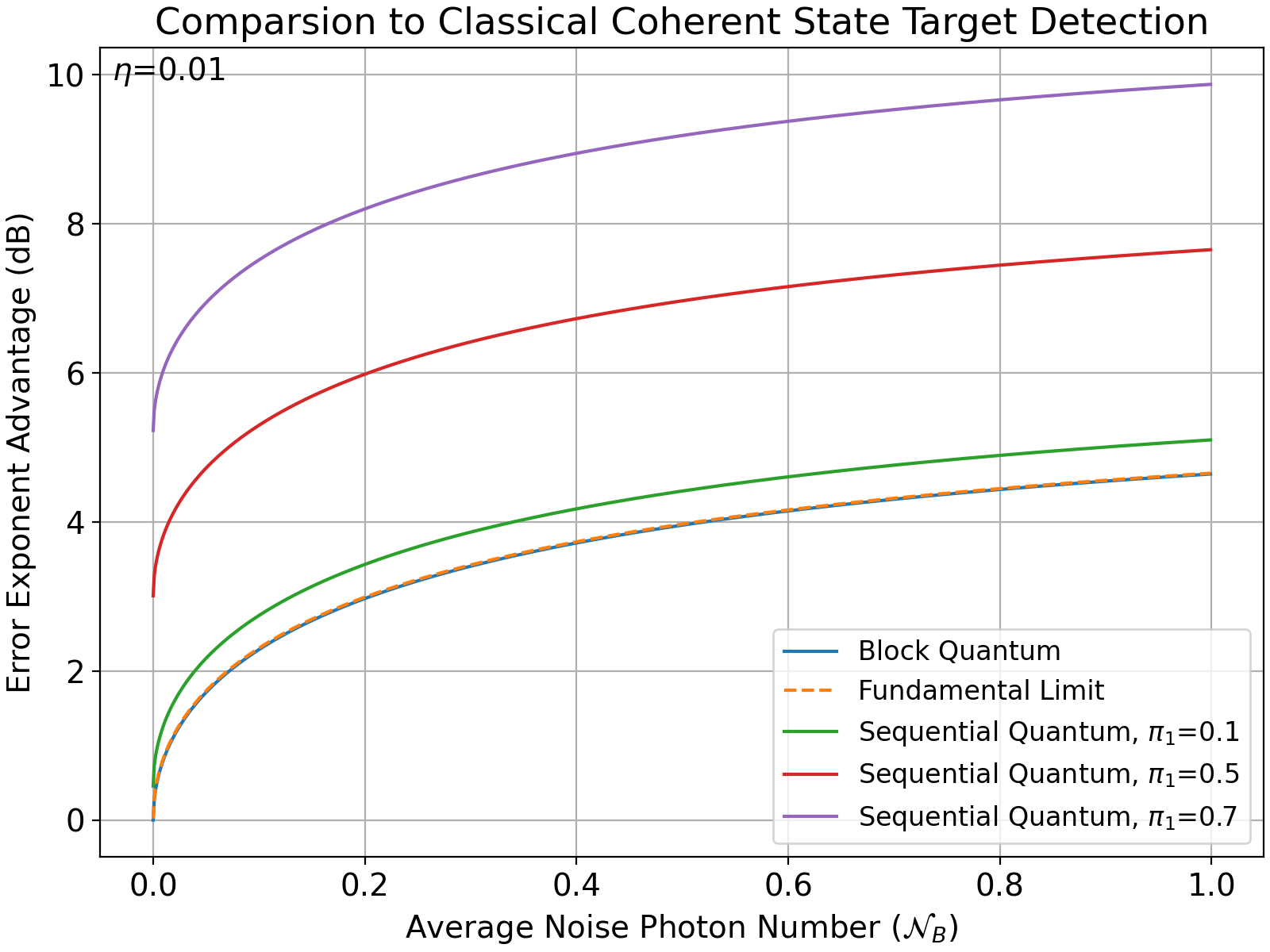}
           \caption{Low Noise}
           \label{fig:advantage_over_coherent_low_noise}
       \end{subfigure}
    ~
       \begin{subfigure}{0.5\textwidth}
           \centering
           \includegraphics[width=0.95\textwidth]{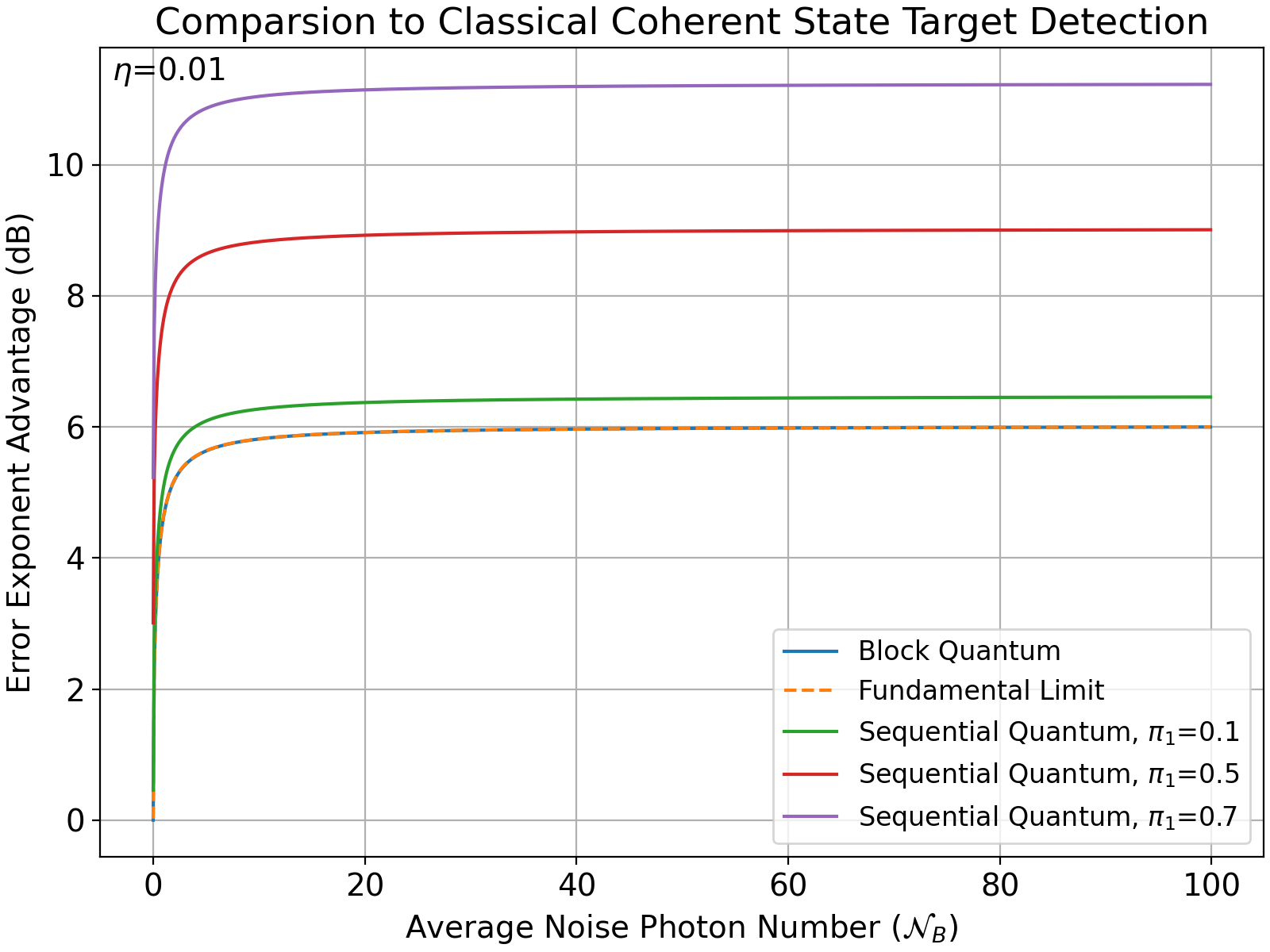}
           \caption{High Noise}
           \label{fig:advantage_over_coherent_high_noise}
       \end{subfigure}
   
       \caption{Comparison of the proposed quantum protocol with both block detection and sequential detection against the coherent state obtained by comparing the error exponent in (\ref{p_e_quantum_proposed}) and (\ref{p_e_quantum_sequential}) to the error exponent in (\ref{coherent_p_e}). The plot also shows the fundamental upper bounds by comparing the error exponent of the coherent state protocol (\ref{coherent_p_e}) to the fundamental limit (\ref{nair_gu_bound}). The fundamental limit perfectly coincides with the proposed block detection scheme.}
       \label{fig:advantage_over_coherent}
   \end{figure*}

   With $N_S$ repeated trials, we have $P_{MD} \leq \left(1 - \tilde{\eta}\right)^{N_S}$ and $P_{FA} \leq \frac{2N_S}{M}$ which in the limit $M \rightarrow \infty$ goes to zero. The performance metric we use is the probability of error, defined as $P_e=\pi_0 P_{FA} + \pi_1 P_{MD} $ where $\pi_0$ and $\pi_1$ are the prior probabilities. To allow for direct comparison to previous work in continuous variable QI, we assume $\langle N_B \rangle_{pm} = \frac{\mathcal{N}_B}{1 - \eta}$ and take the small $\tilde{\eta}$ approximation of $(1 - \tilde{\eta})^{N_S}$ to get the probability of error for the proposed protocol as:
   \begin{align}
    P_e^{QB} & \simeq \pi_1 \exp\left[-\eta N_S\left(\frac{1}{1+\mathcal{N}_B } \right)\right] \label{p_e_quantum_proposed}
   \end{align}
   The optimal classical target detection scheme uses coherent state probes and achieves a probability of error of \cite{tan_et_al}:
   \begin{align}
    P_e^{CS} \simeq \sqrt{\pi_0\pi_1} \exp\left[-\eta N_S \left(\sqrt{\mathcal{N}_B+1}-\sqrt{\mathcal{N}_B}\right)^2 \right] \label{coherent_p_e}
    \end{align}
   Comparing the error exponents in equations (\ref{p_e_quantum_proposed}) and (\ref{coherent_p_e}), we can compute the quantum advantage to be:
   \begin{align}
    \frac{1}{(1+\mathcal{N}_B)\left(\sqrt{\mathcal{N}_B+1}-\sqrt{\mathcal{N}_B}\right)^2 } \xrightarrow{\mathcal{N}_B \rightarrow \infty} 4
   \end{align}
   Therefore, using the proposed QI protocol will always result in a quantum advantage ratio greater than $1$, and this ratio approaches $4$ matching the $6$ dB advantage offered by QI with TMSV in the high-noise regime \cite{tan_et_al}. Figure \ref{fig:advantage_over_coherent_high_noise} shows the quantum advantage and how it scales with increasing noise.
   
   The proposed protocol also extends into the low-noise case. Figure \ref{fig:advantage_over_coherent_low_noise} shows the quantum advantage that can be attained for low-noise quantum illumination between $0 \leq \mathcal{N}_B \leq 1$, increasing from $0$ to roughly $4.6$ dB at $\mathcal{N}_B = 1$. This dispels the commonly held belief that QI has no advantage in the low-noise regime \cite{shapiro_and_lloyd_QI,nair_and_gu}.
   
   Furthermore, in \cite{nair_and_gu}, the transmitter-independent lower bound for quantum illumination in thermal backgrounds is determined to be:
   \begin{align}
    P_e^{QI} \geq \pi_0\pi_1 \exp\left[ N_S \ln\left(1-\frac{\eta}{\mathcal{N}_B+1}\right)\right] \label{nair_gu_bound}
    \end{align}
    The first-order Taylor expansion of the error exponent in this fundamental limit matches the error exponent in (\ref{p_e_quantum_proposed}), indicating that the proposed QI protocol is indeed optimal for small $\eta$ in all ranges of background noise.
   
   We can further improve performance by adopting a sequential decision rule where we transmit photons until we either get a "target present" outcome or reach a predefined threshold $N_{max}$. Since the single-shot probability of a false alarm can be brought to zero, the first positive detection indicates a photon that was reflected by the target, making the sequential decision rule extremely effective with the proposed protocol. With this decision rule, the expected number of transmitted photons is
       $N_{S} = \pi_0 N_{max} + \frac{\pi_1}{\tilde{\eta} }
   $ while the probability of missed detection is $(1 - \tilde{\eta})^{N_{max}}$. In the low $\tilde{\eta}$ approximation and expressing $N_{max}$ in terms of $N_{S}$, we obtain the probability of error:
   \begin{align}
    P_e^{QS} &\sim \exp \left[-\frac{\eta N_S }{(1-\pi_1)(1+\mathcal{N}_B)}\right] \label{p_e_quantum_sequential}
   \end{align}
   Thus, with sequential detection, we obtain an error exponent ratio that is greater than the block quantum case by a factor of $(1 - \pi_1)^{-1}$. Hence, the sequential detection decision rule always outperforms the block decision rule where a constant number of photons are transmitted. The quantum advantage at equal prior probabilities approaches 9 dB when compared to the classical coherent protocol without a sequential decision rule and becomes increasingly advantageous as the prior likelihood of the target being present increases. This advantage is illustrated in Figure \ref{fig:advantage_over_coherent}.
   
   \section{Discussion and Conclusion}
   Prior to this work, the TMSV state was the leading contender for QI. However, the proposed Bell state protocol offers several distinct advantages for practical application. Firstly, both protocols require probing one resolution at a time if the target’s position is unknown. Our preliminary analysis indicates that the Bell state protocol may mitigate this by sequentially performing measurements across all possible resolution bins until a photon is detected. However, this approach increases the rate of missed detection due to repeated measurements on the idler modes. It also increases the likelihood of false alarm because of the increased number of measurements (one per resolution bin). Despite these drawbacks, increasing the dimensionality M is likely enough to overcome the issues and give a quantum advantage. The complete analysis of this enhanced protocol remains the subject of future work and would also address the ranging problem, where the goal is to determine the distance to the target.
   
   Moving towards real-world implementation, both the TMSV and Bell state QI present significant challenges due to current technological limitations in generating the states, storing the idler, and performing the optimal measurement. However, the relevance of Bell states in general quantum applications gives QI with Bell states an advantage, as it benefits from the extensive research base surrounding the engineering of high-dimensional Bell states.
   
   The generation of high-dimensional Bell states in various modes has seen considerable advancements \cite{entaglement_of_orbital_angular_momentum_photons, time_bin_entagled_qubit_by_fs_pulses, generating_distributing_frequency_entagled_qudits, high_dim_spatial_entaglement_photons} and remains an active area of research. While generating arbitrarily high-dimensional Bell states remains challenging, producing the required TMSV states spread across a high time-bandwidth product with significant brightness is arguably even more difficult \cite{QI_review_sorelli_boust}. The rate of convergence to the maximum advantage remains an open problem for both protocols, as each attains maximum advantage only at very high time-bandwidth products as $M \rightarrow \infty$.
   
   Furthermore, any QI protocol necessitates the storage of the idler modes in a quantum memory. Quantum memory capable of storing high-dimensional quantum states is still in its early stages but has been demonstrated in various forms \cite{25d_photonic_qudit_memory, multidim_entaglement_stored_in_crystal, quantum_memory_225_inidividually_memory_cell}. It is important to note, however, that storing DV states is appreciably easier than storing CV states with current technologies \cite{quantum_memories_review}. Using optical delay lines also presents a valid option but significantly limits the range of the QI protocol.
   
   Lastly, the measurement proposed here is novel, and the challenges with its practical implementation have yet to be explored. Since the inception of TMSV QI, there have been several proposals for optimal and non-optimal receivers \cite{optimum_QI_receiver_shapiro, QI_receivers_Guha, QI_reciever_no_joint_meas_zhuang_2024} in efforts to move towards practical implementation. Such receivers are likely to exist for Bell state QI as well since the measurement proposed in this paper resembles a Bell measurement with additional noise photons. Bell measurements represent a crucial class of measurements and have seen several implementation proposals, many of which can potentially be adapted for the measurement proposed in this work \cite{scheme_high_dim_bell_meas, suboptimal_bell_meas_linear}.
   
   In conclusion, we have addressed the formidable challenge of characterizing the performance of the DV quantum illumination scheme based on the M-mode Bell state. Our results indicate that this scheme is optimal across a broad range of noise regimes and noise distributions. While the real-world implementation of quantum illumination remains difficult, our findings challenge existing paradigms regarding the limitations of DV QI, opening the door to exciting new applications and implementations.

   \section*{Acknowledgments}
   We would like to express our gratitude to Han Liu, Zacharie Leger, and Professor Jeffrey H. Shapiro for their helpful discussions and insights.

\bibliography{/Users/arman/documents/bibliography}

\end{document}